# Prediction of Synchrostate Transitions in EEG Signals Using Markov Chain Models

Wasifa Jamal, Saptarshi Das, Ioana-Anastasia Oprescu, and Koushik Maharatna, *Member, IEEE*

***Abstract*****—This paper proposes a stochastic model using the concept of Markov chains for the inter-state transitions of the millisecond order quasi-stable phase synchronized patterns or synchrostates, found in multi-channel Electroencephalogram (EEG) signals. First and second order transition probability matrices are estimated for Markov chain modelling from 100 trials of 128-channel EEG signals during two different face perception tasks. Prediction accuracies with such finite Markov chain models for synchrostate transition are also compared, under a data-partitioning based cross-validation scheme.**

## I. Introduction

PHASE synchronization of multichannel EEG signals has been widely used as a potential measure of brain functional organization and connectivity [1]. Although EEG signals due to their high temporal resolution show highly stochastic temporal evolution, it has been found that the scalp potential topographies are not so random and follow finite sets of small number of quasi-stable patterns which are termed as microstates [2]. Recently, Jamal *et al.* [3] investigated the temporal evolution of the frequency band-specific phase difference topographies to find periods of phase locking in multichannel EEG signals. It has been found in [4] that the phase difference topographies do not change abruptly and microstate-like quasi-stable phase locked patterns are observed in a temporal resolution of the order of milliseconds. These small number of stable phase synchronized patterns are termed as synchrostates, which switches from one to the other within the time interval of a cognitive task. The existence of synchrostates during face perception tasks was first observed in the beta ($\beta$) band (13-30 Hz) with different ensembles of EEG signals [4]. For similar visual stimuli, the inter-state switching patterns only slightly change among different ensembles or trials [4], however it is different for different stimuli and also across different groups of people [3]. Hence, statistical modelling of the pseudo-random and abrupt temporal switching characteristics of synchrostates can be helpful in understanding the dynamic evolution of the stimulus induced brain response particularly in different pathological population. Such a model could be effective in predicting the future behavior of the state transitions in a probabilistic way using a Bayesian like framework, once the initial state is known. Previously, the microstate transitions have also been shown to follow the Markovian property in [5]. In addition, the Markov Chain Monte Carlo (MCMC) approach has been applied to fit neural mass model with EEG signals [6] and for the reconstruction of cortical sources [7]. Studies have shown that in order to mathematically model microstate transitions a higher order or $n$-step Markov model may be needed due to the inherent long range temporal correlations in such sequences [8]. There has been also few attempts to simulate epileptic seizure spikes in EEG using the Markov model and Hidden Markov Model (HMM) [9, 10]. Automated evaluation of stages of sleep from EEG has been modelled using HMM in [11]. Recently, phase synchronization dynamics have been modelled using the HMM and Semi-Markov Model (SMM) in [12] which does not consider the presence of unique phase synchronized states, as done in the present work. The process of deriving synchrostates allows us to represent a multivariate stochastic process (EEG) as a collection of few univariate quasi-static subsystems (unique states) which randomly switches amongst themselves. The unique phase synchronized patterns or synchrostates can be considered as the discrete cognitive states underlying the information exchange and integration within the brain [3]. In contrast to the above mentioned literatures, we first make a probabilistic model of the EEG synchrostate switching sequences using the first and second order Markov chains in order to predict their occurrences and validate the predictions with multiple EEG trials during normal and scrambled face perception tasks.

The present work is aimed to model the switching sequence of synchrostates as a stochastic process over multiple trials, considering that the switching time courses have the Markovian property and hence the source of these switching can be modeled as a finite Markov chain. We used 100 independent trials of EEG signals during scrambled and normal face perception tasks. First order and second order transition probability matrix of Markov chain models were developed using 90% of the data (EEG trials) in order to predict the state transitions from the knowledge of the state at the first time step and the subsequent predictions were verified and compared using the remaining 10% data under a 10-fold cross validation scheme. Markovian property of first and second order inter-synchrostate transition essentially implies

Manuscript received June 27, 2014, Accepted August 15, 2014. The work presented in this paper was supported by FP7 EU funded MICHELANGELO project, Grant Agreement #288241.

W. Jamal, S. Das, I. Oprescu, K. Maharatna are with the School of Electronics and Computer Science, University of Southampton, Southampton SO17 1BJ, United Kingdom (e-mail: {wj4g08, sd2a11, io1g10, km3}@ecs.soton.ac.uk).

that the value of each state at any time instant depends only on the state in the last one/two previous step(s) respectively.

## II. Theoretical Background

### A. Synchrostate Sequences from Multichannel EEG

To derive the synchrostates [3], multichannel EEG signals $x_i(t)$ where $i$ is the electrode, undergo complex Morlet wavelet transform that produces the instantaneous phase of the signal $\varphi_i(a,t)$, captured at each electrode as a function of scale $a$ (frequency) and time $t$. Using the instantaneous phases a time evolving phase difference matrix $\Delta\varphi_{ij}(a,t)=|\varphi_i(a,t)-\varphi_j(a,t)|$ corresponding to a band of frequency (or wavelet scale) can be constructed. These matrices when clustered using the unsupervised $k$-means clustering algorithm yield synchrostates in which there is little variation in their phase difference topography [3, 4, 13]. The optimal number of clusters underlying a dataset is determined by an incremental $k$-means clustering algorithm. Typically, the number of synchrostates are found to be three to seven and their presence are shown during the execution of face perceptions tasks in adults, typical and autistic children [3, 4]. During the quasi-static period of the synchrostates, the phase topographies of the EEG signals have no significant variation and hence can be considered to be in synchrony. The synchrostates exhibit switching patterns among themselves and the temporal characteristics of such switching mostly depend on the particular cognitive task. This dynamic inter-synchrostate switching process is modelled here in a probabilistic fashion using finite Markov chains.

### B. Markov Chain Models for Inter-synchrostate Transitions

The probabilistic evolution of many dynamical systems has been modeled by Markov chains [14]. The Markov chain can jump from one state or condition to another, provided the transition is probabilistic and not deterministic. Due to the probabilistic nature of the model it cannot predict the future states from the present with certainty, however it can assign probabilities to the possible states that can occur. Thus in a Markov process the future states are assessed by a vector of probabilities [15]. The evolution of these vectors essentially describes the underlying dynamical nature of a system. In a first order Markov chain, the state at any time instant depends only on the state immediately preceding it, and hence is defined as a single-dependence chain. However, in Markov chains with higher dependency relationships like second or higher order chains, the subsequent state depends on two or more preceding ones.

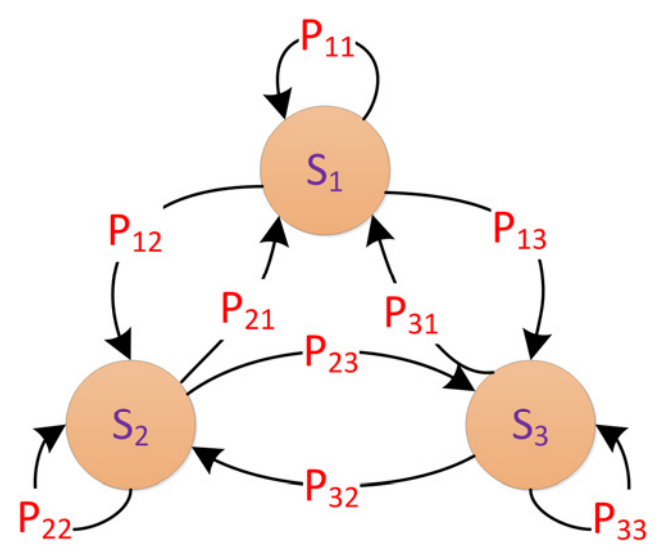

Fig. 1. The state transition diagram for three synchrostates.

In an $n^{th}$ order discrete Markov chain, the process can be in any one of the finite number ($m$) of possible states $\{S_1,S_2,\ldots,S_m\}$ at any time instant. As the chain progresses, the states may change from one to another. This process is determined by transition probabilities between discrete states in the observed system which is estimated using the maximum likelihood approach [16], where $p_{ij} = N_{ij} \Big/ \sum_j N_{ij}$, $i=1,2,..,m,$ $j=1,2,\ldots,m$. Here, $N_{ij}$ is the number of transitions from state $i$ to $j$. Given an initial condition (state), if the process is in $S_i$ at time $n$, then at time $(n+1)$ it will be at state $S_j$ with probability $p_{ij}$. In this study, we only consider stationary Markov chains i.e. $p_{ij}$ does not vary with time or space [15, 16]. The transition probabilities, $p_{ij}$ of Markov chain are considered as the elements of the $m\times m$ non-negative stochastic matrix $P$, commonly known as the state transition matrix. The sum of the transition probabilities along each row of the transition matrix $P$ equals one. If we look at the Markov process after two steps given an initial state $S_i$, the transition is governed by applying the underlying transition matrix, $P$ twice. In other words if $p_{ij}^{(2)}$ is the transition probability of reaching state $S_j$ from initial state $S_i$ in two steps, then $p_{ij}^{(2)} = \sum_{k=1}^{m} p_{ik}p_{kj} = \left[P^2\right]_{ij}$.

Therefore, the two-step transition matrix is given by $P^2$, the three step transition matrix is given by $P^3$ and $n$-step transition matrix is $P^n$, such that the $ij^{th}$ entry of $P^n$ is the probability of the system reaching state $i$ to state $j$ in $n$ steps.

The basic limit theorem [14] states that for certain types of Markov chains there exists a unique limiting probability vector $p^T$. In other words, in $n$-steps for any initial state $i$ the transition matrix tends toward a limit $m\times m$ matrix, $\overline{P}$, known as the steady state transition matrix, each of whose rows equals $p^T$ i.e. $\lim_{n\to\infty} P^n = \overline{P}$, where each row of $P^n$ converges to $p^T$, as $n\to\infty$. This types of chains are called regular Markov chains. A Markov chain can be considered as a linear dynamical system with a positive system matrix [14]. We show a schematic representation of the transition amongst three synchrostates as an example case in Fig. 1 where $S_i$ represents state $i$ and $p_{ij}$, $i,j=\{1,2,3\}$ indicates the probabilities of switching from state $i$ to $j$ which needs to be estimated from the observed synchrostate sequence dataset. Once the transition probability matrix is obtained, it is possible to predict the future steps of the synchrostate transition given an initial state using the first and second order Markov chain models [16, 17]. Although higher order (>2) Markov chain models may give better results for prediction they are prone to over-fit the training data. Hence we restricted our study to first and second order Markov chain models.

## III. Simulation and Results

The synchrostates analysis was carried out on the SPM multimodal face-evoked dataset [18]. The dataset consisted of 128-channel EEG signals acquired from an adult during the execution of face perception tasks when presented with multiple normal and scrambled face stimuli. The 100 trials of EEG signals were epoched and pre-processed and then different ensembles of the data were segmented into 10 equal partitions, each of them containing 10 trials of the EEG. The phase response of each individual 10 segments of EEG for both scrambled and normal face stimulus, were clustered using incremental *k*-means clustering algorithm following the technique proposed in [3] to obtain the synchrostates. The clustering results of EEG in the $\beta$ band have yielded optimal three synchrostates for both the face stimuli for all the ten segments. The clustering also generated associated inter-synchrostate switching sequence patterns which may be described as probabilistic switching between the three discrete and unique synchrostates in a configuration of Fig. 1, over the task completion time of 400 time steps. The temporal switching patterns amongst these states during normal face stimuli were found to be similar across different ensemble of trials [4], however they differ between two stimuli (normal and scrambled face) and thus could be considered as a unique signature of the visual stimuli provided. This allows us to generate two probabilistic models of first and second order Markov chain to fit the state transition dynamics for each of the two stimuli.

The state transition sequences for the whole 100 trials without data-partitioning have been shown in Fig. 2 for both the normal and scrambled face stimuli. The associated optimal three synchrostate topographies have also been depicted in Fig. 2 for the whole 100 trials of EEG taken together. It is evident from Fig. 2 that the state topographies are almost similar for both the stimuli but their transition sequences differ significantly. For example during normal face perception the sequence starts from state 3 whereas for scrambled face perception it starts from state 2. In addition, for the normal face perception state 2 occurs minimum times whereas for scrambled face perception state 1 occurs the least times, indicating the cognitive task-specific nature of the synchrostate switching patterns. The principle diagonal elements of the transition probability matrix $p_{ii}$, $i=1,2,3$ can now easily be estimated from the sequences shown in Fig. 2 with prevalence of the same state and so as for the rest of the terms $p_{ij}$, $i \neq j$, by counting the number of transitions. From the switching sequences obtained for each of the $k=10$ folds of the partitioned EEG trials, synchrostate switching patterns are derived next yielding a similar characteristics like in Fig. 2.

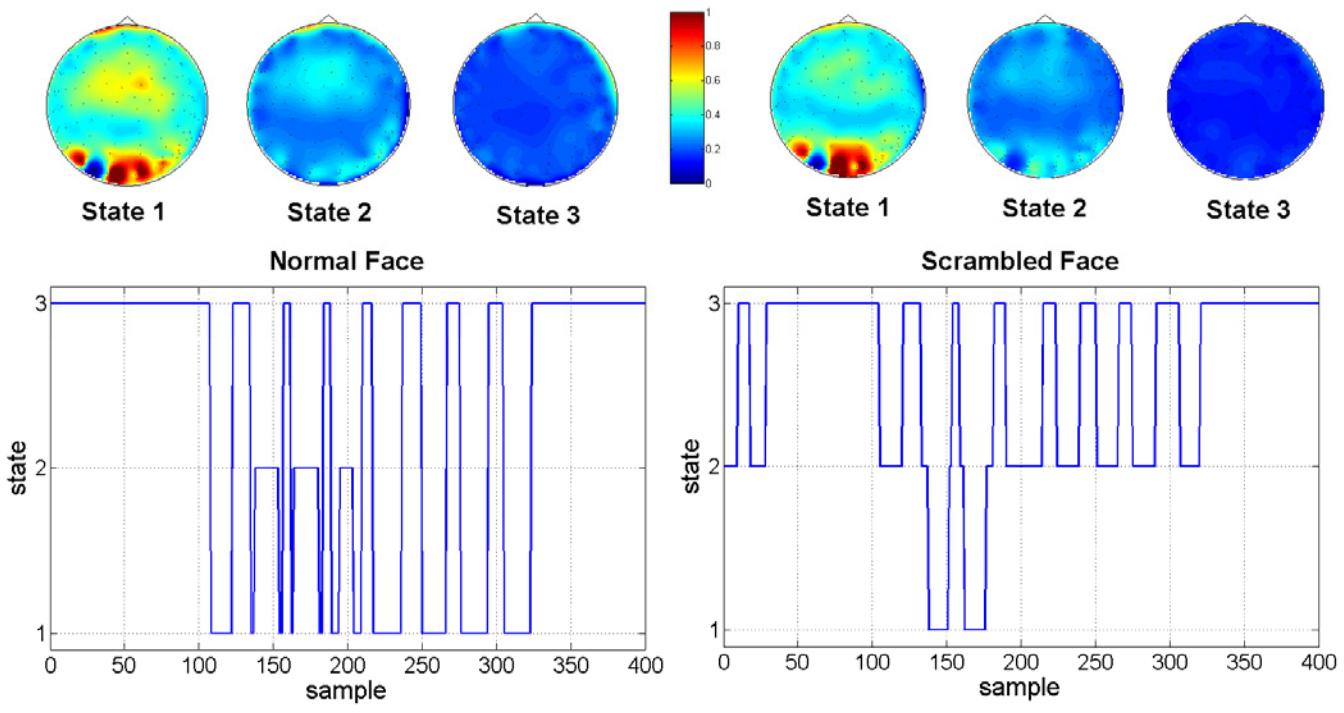

Fig. 2. Synchrostate topographies and state transitions for normal and scrambled face perception for 100 EEG trials.

We now aim to model and simulate the switching sequences of synchrostates as a finite Markov process for each of the $k=10$ folds of synchrostate switching diagrams based on the characteristics of 90 EEG signals. Starting from the ten group (or fold) of EEG synchrostate observations, a cross-validation scheme has been adopted to generalize the model across different ensembles (or group of trials) and generate the probabilistic model which can give best use of limited data with less chance of introducing bias from the validation data-set [19]. During our experiment, each of the single folds containing 10 EEG trials was held out as the validation dataset and then the rest 9 folds containing 90 EEG signals were used to train the probabilistic model.

We use the limit theorem to consider the long term performance of our estimated model. Fig. 3 shows that the synchrostate transition is a regular Markov chain process when estimated on the whole 100 trials of the data. This has been verified by obtaining the state transition matrix $P$ and then raising the power as $P^n$ as $n\rightarrow\infty$. Fig. 3 shows that all the 9 elements of the transition matrix obtained from the three synchrostate switching sequences converges to the steady state transition probability or eigen-vectors of the state transition matrix as the number of time steps are increased [14, 15]. The steady state probabilities of the three synchrostates are found to be $p^T_{normal}$ = {0.6763, 0.13, 0.1937} for normal face stimulus and $p^T_{scrambled}$ = {0.6301, 0.2504, 0.1195} for the scrambled face stimulus respectively, no matter at which state the sequence or chain has started.

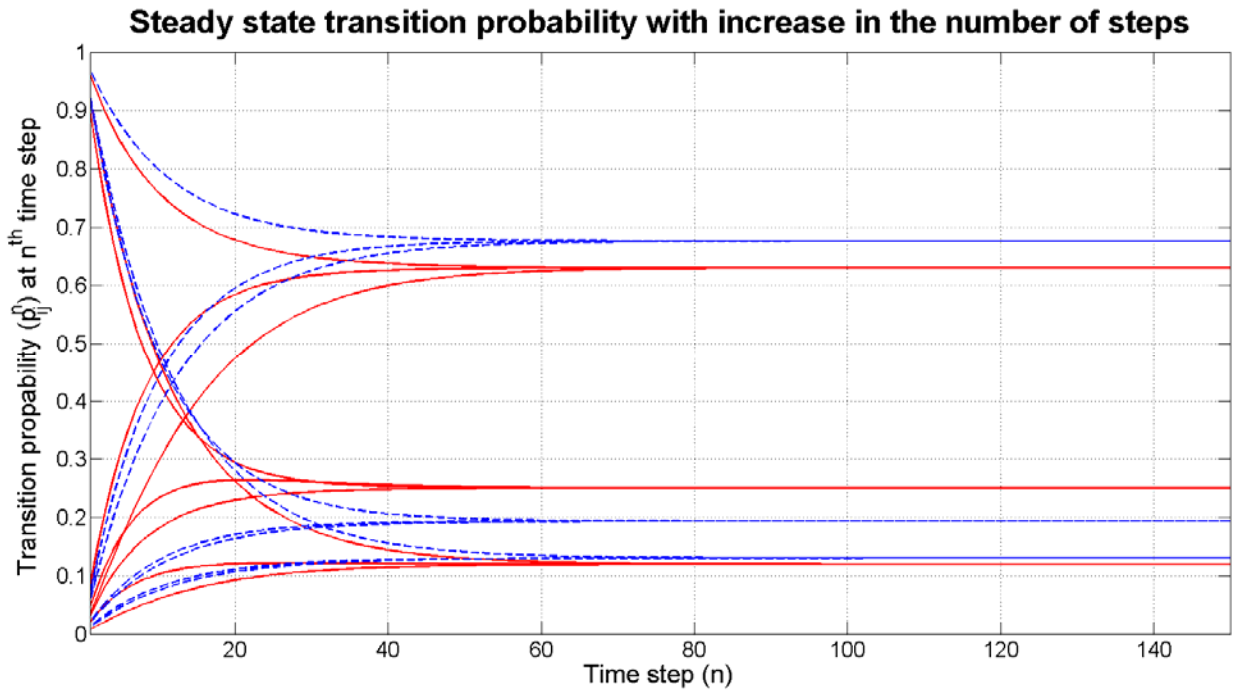

Fig. 3. Steady state transition probability for normal and scrambled face stimuli (continuous line (red) – scrambled, dashed line (blue) – normal face).

For the validation of the Markov model, the synthetic generation of state sequences is simple and straightforward once the model is built from the 9-folds of the whole dataset. From the estimated Markov model representing a stochastic dynamical system, the outcome as the synchrostate switching sequences will vary in different realizations of the underlying random process, due to the probabilistic nature of the problem. Therefore, during the validation phase, the synthetic data from the same Markov chain model will be different considering multiple independent realizations of the same Markov chain given the initial synchrostate condition at the beginning of the cognitive task. Also, it is mathematically incorrect to match a real data with a single outcome of a trained Markov process. To circumvent this problem, within each fold of data and at each time step, by referring to the estimated transition matrix and given initial state, the program makes 100 independent realizations for the prediction of which state the system moves to in subsequent time steps, using a discrete random sequence generator [20]. Based on the estimated or trained Markov model using the past $n$ number of samples, the expected value of 100 independent predictions of the possible state at step $(n+1)$ have been validated with the real observation of the held out state at time step $(n+1)$. The mis-predictions are tracked over all the 100 independent realizations and across the 400 time steps for all the $k=10$ folds of data segments. The misprediction rate or error for each fold is then averaged to produce the average error rate of the model for a particular order (first or second) of Markov chain. For building the second order Markov chain model the $(n+2)^{th}$ sample has been predicted in a similar way given the synchrostate knowledge at time steps $n$ and $(n+1)$. The Markov chain model training and validation algorithm for the synchrostate transition is illustrated in the following steps:

**Step1:** For each fold $i$, $i=\{1,\ldots,k\}$, calculate the transition probability matrix $P$ by taking the average of all the $P$'s over the 90% of the all training sequences.

**Step2:** Given the knowledge of the initial state from the test sequence generate the discrete events for the next time step for 100 independent realizations. This produces random states from the discrete probability values of the state transition matrix, using Matlab function *gendist* [20].

**Step3:** Compare the 100 predicted states with the observed state in the test sequence. If mispredicted, increase the error counter.

**Step4:** Increment iterations for the next time step and repeat steps 2 to 3.

**Step5:** Compute the expected error across the 100 independent realizations of the Markov model.

Applying the above proposed algorithm yields Fig. 4 which shows that the error rates for each of the first and second order Markov models for normal and scrambled face stimuli. The median percentage errors for the first and second order Markov chains for normal and scrambled face are 8.49, 8.37, 10.68 and 10.7 respectively. The small median value and inter-quartile ranges of the error rates for the two first order Markov chain models indicates that the model is quite successful in predicting the synchrostate transitions. In the present study, the normal face perception related Markov model performs better than the scrambled face one, as evident from the smaller interquartile ranges as well as the medians in Fig. 4.

Also, given the state transition matrix $P^n$, it is possible to compute the probability of getting state $j$ starting from state $i$ in $n$ time steps i.e. $S_i(n) = S_i(0)P^n$. This allows us to check the Markovian property of the data using the estimated model for predicting the state at $n^{th}$ time-step from the knowledge of the initial state. We ran simulations for $n$=400 subsequent time steps and plotted the prediction errors for 100 different realizations over all the 10-folds as shown in Fig. 5. It is evident that the long-term prediction from a given initial state becomes poorer as the error bounds diverges and becomes more prone to outliers as time evolves.

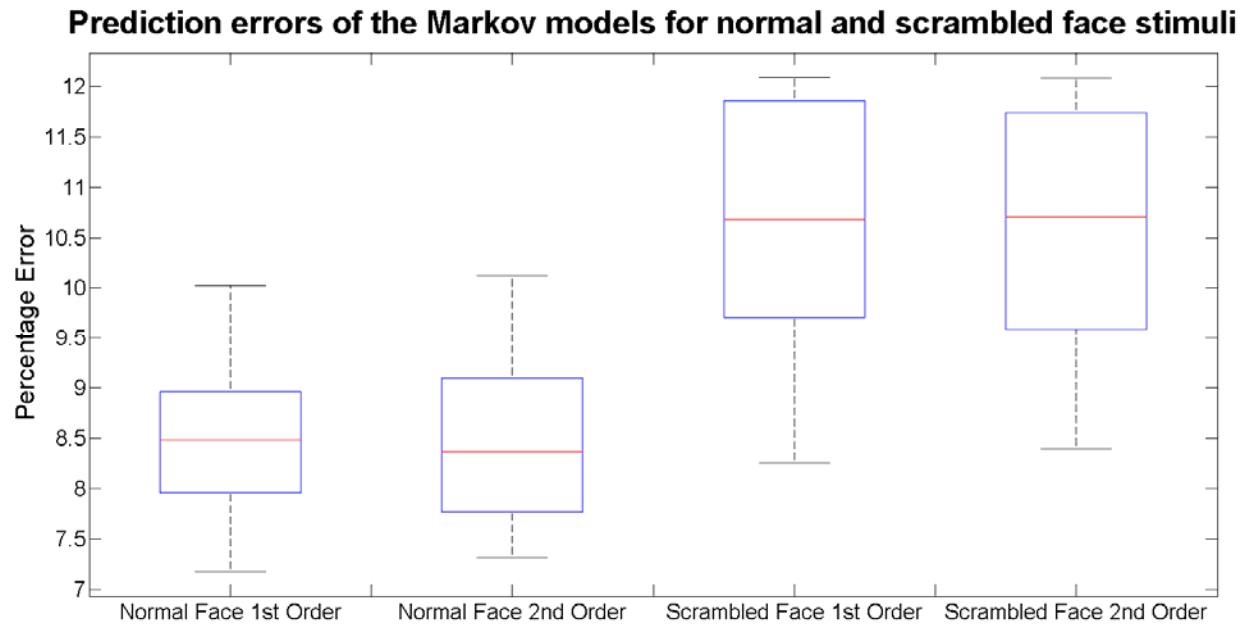

Fig. 4. Average prediction errors using 1st and 2nd order Markov model.

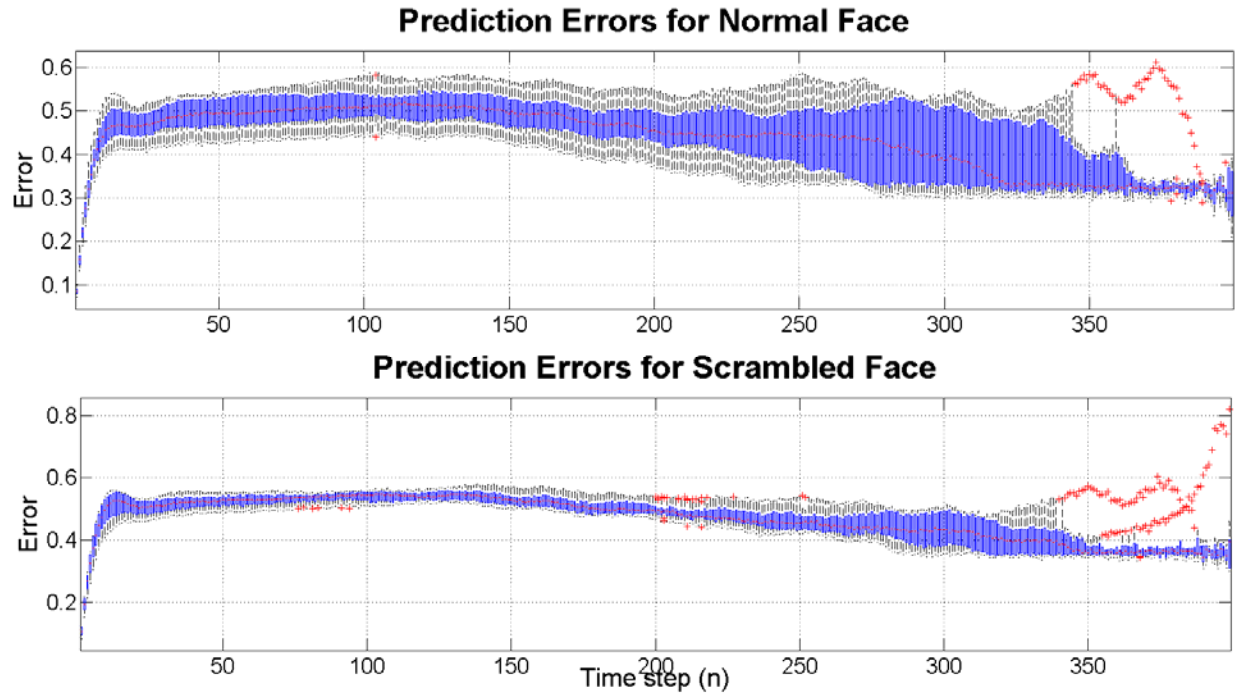

Fig. 5. Box-plot of the error rates over all the 10-folds across the $n$ time steps.

## IV. Conclusion

In this study, a probabilistic model is developed in order to synthetically generate the EEG synchrostate switching sequences as first and second order Markov process and then to validate the predictions using a 10-fold cross validation scheme. The Markov model provides interesting information about the temporal evolution process of the synchrostates characterizing the underlying probabilistic brain dynamics. Our probabilistic model successfully predicts the inter-synchrostate switching patterns with the best average accuracy of 91.63% (for normal face perception) and 89.32% (for scrambled face perception). The proposed modeling approach may shed new light in understanding the stochastic dynamical basis of cognition in humans and prediction of the semi-deterministic switching behavior within the discrete set of phase-synchronized patterns or synchrostates.